# Secret Sharing and Proactive Renewal of Shares in Hierarchical Groups

Ruchira Naskar

School of Information Technology

Indian Institute of Technology, Kharagpur

`Ruchira.Naskar@sit.iitkgp.ernet.in`

Indranil Sengupta

Department of Computer Science and Engineering

Indian Institute of Technology, Kharagpur

`isg@iitkgp.ac.in`

## Abstract

Secret sharing in user hierarchy represents a challenging area for research. Although a lot of work has already been done in this direction, this paper presents a novel approach to share a secret among a hierarchy of users while overcoming the limitations of the already existing mechanisms. Our work is based on traditional $(k+1, n)$-threshold secret sharing, which is secure as long as an adversary can compromise not more than $k$ secret shares. But in real life it is often feasible for an adversary to obtain more than $k$ shares over a long period of time. So, in our work we also present a way to overcome this vulnerability, while implementing our hierarchical secret sharing scheme. The use of Elliptic Curve Cryptography makes the computations easier and faster in our work.

## 1  Introduction

The need to maintain the confidentiality of any data within an organization as well as to prevent it from becoming corrupted or inaccessible due to single point failure, has grown considerably, with rapid increase in computer crime. To provide a solution to this, a lot of research work has already been done to share a secret among a group of users. Parallel to this, many present day applications arrange such groups of users into a hierarchy. For example, in a banking system, the employees are arranged into a hierarchy according to their ranks or designations. In this paper we propose a scheme to share a secret among all the users forming such a hierarchy. Our scheme is based on traditional $(k+1, n)$-threshold secret sharing (where a secret data is divided into $n$ shares and stored into $n$ different locations).

An important observation regarding traditional $(k+1, n)$-threshold secret sharing is that, it is secure as long as an adversary can compromise not more than $k$ secret shares, whereas in real life it is often feasible for a mobile adversary to obtain more than $k$ shares over entire lifetime of the scheme. That is why proactive secret sharing scheme is developed which divides the entire lifetime of the scheme into several time periods and in each time period the shares held by each user is renewed. So, the secret shares obtained from various locations by the adversary in one time period are rendered useless in the next time period. In this paper we show how a secret can be proactively shared into a hierarchy. That is, we add to our hierarchical secret sharing scheme, features, which renew the secret shares held by the members of the hierarchy periodically.

Before proceeding further, we shall explain the aim of this paper clearly. To do so we need to describe and to analyze two well known hierarchical secret sharing schemes, which is done below.

**Shamir's Hierarchical Secret Sharing** Shamir [1] in his hierarchical secret sharing scheme proposed to assign larger number of shares to users at higher levels of the hierarchy so that higher level users hold larger shares of the secret and lower level users hold smaller shares. This leads to the fact that to reconstruct the secret in Shamir's scheme, the number of lower level users required is greater than number of higher level users. As a result there remains no distinction among the secrets held by the users at different levels (i.e. a number of lower level users are equivalent to one higher level user). At the same time the higher level users face storage problem since they need to store a large number of secret shares in this scheme.

**Tassa's Hierarchical Secret Sharing** Tassa [2] introduced the concept of qualitative distinction among the users in their hierarchical secret sharing scheme, by the usage of polynomial derivatives, to generate lesser shares for lower level users instead of representing the secret as the free co-efficient of a polynomial as in Shamir's scheme. The secret is reconstructed in this scheme by applying Birkhoff interpolation. Tassa's scheme suffers the limitation that use of a large finite field is necessary since the size of the base field gets reduced each time a new level is added to the hierarchy and the polynomial is differentiated.

Our goal in this paper is threefold. (1) To propose an efficient hierarchical secret sharing scheme which eliminates the storage problem of Shamir's hierarchical secret sharing scheme. (2) To ensure that the proposed scheme does not reduce the size of the base field, as we go down to the lower lev-els of the hierarchy, while sharing a secret, thus eliminating the limitation of Tassa's scheme. (3) To propose a proactive share renewal mechanism that can be applied to our hierarchical secret sharing scheme which makes it secure in an environment, where a mobile adversary may be present throughout the entire lifetime of the scheme.

## 1.1  The Hierarchical Model

To implement secret sharing into a hierarchy, we partition a given hierarchy into some levels, such that the more powerful members form the higher levels and less powerful ones form the lower levels of the hierarchy. We refer to the power possessed by the members, in the context of what fraction of the secret is held by each of them. More powerful members, comprising the higher levels of the hierarchy hold larger fractions of the secret, while the less powerful members, comprising the lower levels hold smaller fractions. We refer to the above mentioned hierarchy, more appropriately, as a tree throughout this paper.

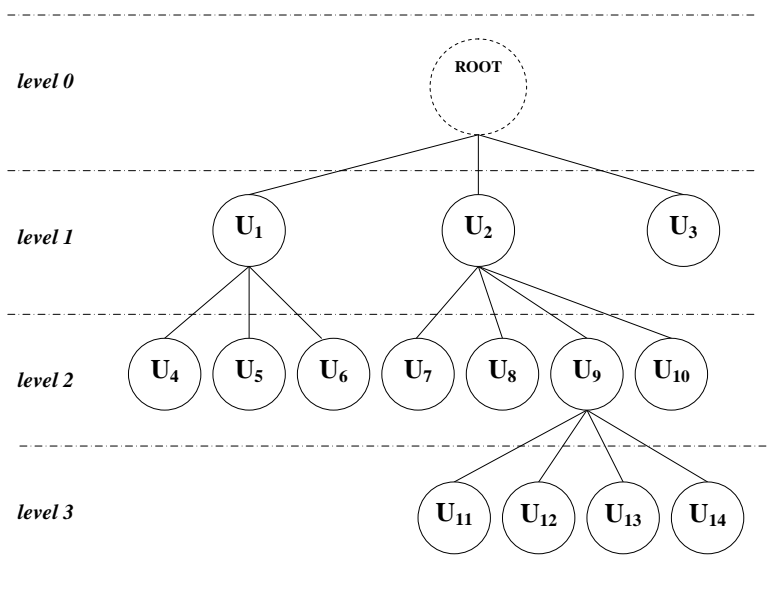

Figure 1: User Hierarchy.

The idea of hierarchical arrangement of users is made clear by figure 1. The figure shows that the users are arranged into a hierarchical tree (according to their positions in an organization). The tree is divided into a number of levels. There is a trusted entity or a server forming the root of the tree. The root server is responsible for secret sharing, delegation and reconstruction. All the users or members of the hierarchy are arranged into the lower levels.

In our scheme, the secret is divided into several pieces and those pieces are distributed among all the users forming the hierarchy in such a way that with each increasing level, the amount of information about the secret, contained in the delegated share, decreases. We have introduced the concept of threshold factor and split operation in this paper, and have used elliptic curve cryptography in order to make the computations faster and easier. Security of our work depends on the difficulty of solving ECDLP (Elliptic Curve Discrete Logarithm Problem).

**Organisation of the paper.** Section 2 presents a review of related research works. Our hierarchical secret sharing scheme has been presented in details in section 3. In section 3.6 an analysis is made, on how far the proposed hierarchical secret sharing scheme has succeeded, to overcome the limitations of the existing schemes. The proactive share renewal mechanism in a hierarchy has been described in section 4. Section 4.1 gives a security model for the proactive share renewal, section 4.3 presents a share renewal protocol and section 4.4 presents a protocol to detect a compromised node. We analyze the efficiency and performance of the proposed protocols as well as compare our work with existing schemes in chapter 5. Finally we conclude the paper in section 6.

## 2 Review of Literature

In this chapter we first present the traditional threshold secret sharing scheme proposed by Shamir [1]. Having discussed the traditional threshold secret sharing scheme, we describe hierarchical threshold secret sharing schemes of Shamir [1] and Tassa [2], along with some other remarkable research works on hierarchical access structures and access rights. Finally we present some well known proactive secret sharing schemes.

### 2.1 Traditional Threshold Secret Sharing

In the original paper by Shamir [1], an elegant way to share a secret data $D$ into a group of $n$ users has been showed. The secret data $D$ is shared among $n$ users in such a way that it can be reconstructed easily from any $(k + 1)$ or more shares. But the knowledge of $k$ or less shares gives no information about $D$. This secret sharing scheme is called the $(k + 1,n)$ threshold secret sharing scheme, where $n$ is the number of users among whom the secret is shared and $(k + 1)$ is the threshold, i.e. the least number of usersrequired to reconstruct the secret.

Thus, the $(k + 1,n)$ threshold secret sharing scheme was originally proposed by Shamir in 1979. In his paper, Shamir [1] stated the goal of a

$(k + 1, n)$-threshold secret sharing scheme to be:
Dividing a secret data $D$ into $n$ pieces in such a way that:

1. knowledge of any $k + 1$ or more $D_i$ pieces makes $D$ easily computable;

2. knowledge of any $k$ or fewer $D_i$ pieces leaves $D$ completely undetermined (in the sense that all its possible values are equally likely).

To each of the $n$ users, one point in a two-dimensional plane is allocated, represented as, $(x_1, y_1),...,(x_n, y_n)$. Each $x_i$ is distinct and $y_i = q(x_i)$ for all $i$ where $q(x_i)$ is a random $k$ degree polynomial of the form $q(x_i) = a_0 + a_1 x_i + ...a_k x_i^k$ in which $a_0 = D$ and $a_1...a_k$ are random integers. The secret share for $i$th user is computed as $D_i = q(x_i)$. Knowledge of any $k + 1$ of those shares, is sufficient to find $a_0...a_k$ by interpolation, where $a_0$ represents the original secret.

By Shamir's [1] scheme, if a secret is shared among a group of users, not all of the users are required to co-operate, which ensures convenience of use. On the other hand, if a number of users less than a threshold ($k + 1$ here), co-operate, they can never reconstruct the secret, at least a number of users, greater than the threshold, is required for the same. This ensures reliability of the scheme.

## 2.2 Hierarchical Secret Sharing Schemes

### 2.2.1 Adi Shamir [1]

Shamir proposed a way to apply his $(k + 1, n)$ threshold secret sharing scheme [1] to a hierarchical structure. To do so, he proposed to grant larger number of secret shares to more powerful users (forming the higher levels of hierarchy) while less powerful users, at the lower levels of the hierarchy are granted lesser number of shares. This leads to the fact that lesser number of higher level users are capable of reconstructing the secret whereas the same action requires more number of lower level users.

We give an example to make the idea clear. Consider a hierarchy of 7 users, consisting of three levels 0, 1 and 2. Following Shamir's hierarchical secret sharing scheme, the user(s) at *level* 0 are given 3 shares each, users at *level* 1 are given 2 shares each and users at *level* 2 are given 1 share each. Now, consider a scenario of (3,n) threshold secret sharing applied to this hierarchy. (Here $n$ is 7 and $k$ is 2.) From the working of Shamir's $(k + 1, n)$ threshold secret sharing scheme and from the share distribution in the hierarchy, it is clear that to reconstruct the secret only 1 *level* 0 user is sufficient, whereas it requires at least 2 users at *level* 1 and at least 3 users at *level* 2.

Clearly, Shamir's [1] hierarchical secret sharing scheme suffers from storage problem since the higher level users have to store larger number of shares.

### 2.2.2 Tamir Tassa [2]

In 2007, Tamir Tassa [2] introduced the concept of distinguishing the hierarchical levels qualitatively, i.e. the secret share of a higher level user contains more information about the original secret than that of a lower level user. Thus he solved the problem of large storage requirement faced by Shamir's hierarchical secret sharing scheme. Tassa [2] used derivatives of polynomials instead of using free coefficients of polynomials to represent the secret.

With each lowering level in the hierarchy, we find that the size of the base field gets reduced by this scheme, although it solves the storage problem faced by Shamir's [1] scheme. To reconstruct the secret, Birkhoff interpolation is used in Tassa's [2] scheme which is applicable to a set of polynomial derivatives.

In addition to this, a lot of research work have been done on access structures and access rights in user hierarchy. Among the notable ones, is a delegation mechanism, proposed in 1997 by Chris Charnes et. al. [14], where a higher level user in a hierarchy can delegate his power to retrieve a secret, to some lower level user. The above mentioned paper deals with access structure in a hierarchy and authorized subset of users. In 2004 Chang et. al. dealt with access structure in user hierarchy and delegation of access rights in [15], where each access structure has its own secret key and one class of user can delegate its access right to another class by granting admission ticket to it.

## 2.3 Proactive Secret Sharing

Traditional $(k+1,n)$ secret sharing schemes share a secret among $n$ users in such a way that any $k+1$ of them can reconstruct the secret, but $k$ or less shares are incapable to do the same. This assumes that an adversary can never gain more than $k$ shares. But in an environment where an adversary is mobile and can compromise more than $k$ users, such schemes fail. So, proactive secret sharing schemes first share a secret among the users, using a traditional secret sharing scheme, and then divide the entire lifetime of the scheme into a number of time periods (a day, a week, or a month) in such a way, that no adversary can gain more than $k$ shares during a single time period. During each period, the shares of the users are renewed, so that the shares obtained by an adversary during one time period are rendered useless from the next time period.

### 2.3.1 Otrovsky and Yung [5]

In order to keep servers free from mobile adversaries such as network viruses or worms, Otrovsky and Yung [5], in 1991, proposed the first proactive secret sharing scheme, where old shares are discarded and new shares are generated periodically and a server, compromised by an adversary, is rebooted to get rid of the adversary.

### 2.3.2 Herzberg et. al. [3]

In 1995, Herzberg et. al. [3] proposed a proactive secret sharing scheme which renews the shares of all users forming a group periodically, so that, any share compromised by an adversary during one time period, becomes useless from the next. In Herzberg et. al.'s [3] scheme, initially all the users obtain their secret shares by any traditional threshold secret sharing. To renew a secret share, Herzberg et. al. [3] proposed to add to the share, a set of polynomials, having their free co-efficients equal to 0 (i.e. $\delta(0) = 0$ where $p$ is the polynomial). Each such polynomial is sent to any user, by one of the other users. That is, in time period $t + 1$, user $U_i$ renews his share $q^{(t)}(i)$ to $q^{(t+1)}(i)$ by computing $q^{(t+1)}(i) = q^{(t)}(i) + \sum_{k=1}^{n} \delta_k(i)$, $\forall$ k such that $U_k$ belongs to the group (of $n$ users) and where $\delta_k(i)$ is the polynomial contributed by user $U_k$. Thus, Herzberg et. al.'s [3] scheme depends on all users in a group to renew the share of any one user. To find out whether any adversary is present within the group, or any node has become compromised, the proposed scheme uses a verification mechanism which depends on Discrete Logarithm Problem.

### 2.3.3 David A. Schultz [6]

Another remarkable research work in this direction is that of David Andrew Schultz [6]. In 2007, Schultz [6], in his thesis on "'Mobile Proactive Secret Sharing"', proposed an extension of proactive secret sharing scheme, which allows the group members participating in the scheme to change dynamically from one time period to another. In addition to this, in Schultz's [6] scheme the threshold is also variable between different time periods. Thus Schultz's [6] mobile proactive secret sahring scheme is much more flexible that the previous proactive secret sharing schemes.

## 3   Hierarchical Secret Sharing

In this section we describe in details, the proposed way of sharing a secret into a hierarchy. Before moving into the details it is important to mention clearly all the assumptions we have made in this scheme, which are:

1. Each user or member of the hierarchy, knows his position in the hierarchy, i.e. who is his immediate predecessor and who all are his immediate successors in the tree.

2. The presence of a trusted entity in the organization, which we refer to as the *root server*.

3. To participate in the proposed hierarchical secret sharing scheme, each user has to register himself with the server, through which he is granted a unique secret *registration token* which is known only to that user and the server. A user obtains his registration token from the server either over a secure channel (e.g. in person or by courier). Registration is done only once by each user when he joins the group (of users forming the hierarchy) for the first time.

### 3.1 Important Definitions and Notations

Our scheme is a threshold secret sharing scheme and we have introduced the concept of *threshold factor* in this paper.

**Definition 1.** The *threshold factor* $(TF)$ is a positive integer whose range is (0,1]. If a secret is shared among $n$ number of users, the minimum number of them required to co-operate for reconstructing the secret correctly, is the ceiling of the product of $TF$ and $n$. That is, in terms of $(k, n)$ threshold secret sharing,

$$\text{threshold}(k) = \lceil TF \ . \ n \rceil.$$

For example, if a secret is shared among 9 users and the $TF$ is 0.3, then the number of users (from those 9 users) required to reconstruct the secret will be threshold$(k) = \lceil 0.3 \times 9 \rceil = 3$.

*Split operation* is a basic operation used to implement the proposed scheme. It is defined as follows.

**Definition 2.** Given a positive integer $I$, two random positive integers $I'$ and $I''$ are chosen such that $I' + I' = I$. This is called the *split* operation.

Next we present some important notations:

**E**$(F_p)$ **-** An elliptic curve modulo $p$ on prime finite field $F_p$ of size $p$, having a base point $G$ which is made available publicly to all users in the hierarchy.

$rtok_i$ **-** *Registration token of user $U_i$.*
Obtained from the server by $U_i$ very securely, preferably in person or by courier, during registration while joining the group.

$K_i^g$ **-** *Group key of user $U_i$.*
Maintained and used as long as $U_i$ is a member of the group.

$K_i^r$ - *Round key of user $U_i$.*
> Maintained and used for one round of secret distribution and reconstruction.

$P_s^r$ - *Server's Public Round Key.*
> Maintained and used for one round of secret distribution and reconstruction. Used by the group members to compute their round keys.

## 3.2   Joining the Hierarchy by a User

(a) *Registration:* When a new user wants to join an existing group of users forming the hierarchy, he has to make a registration with the server. While registration the user obtains a unique registration token from the server (as described earlier), using which he computes his group key. Each user maintains this registration token and the group key as long as he is a member of the group. The server also stores the group key of each group member as long as he is a member of the group.

An example will make the idea clear. Let $U_i$ be a user who wants to join an already existing group of users. $U_i$ requests the server for registration. On receiving this request, the server grants $U_i$ a registration token $rtok_i$, which is stored by $U_i$ as long as he is a member of the group. Registration token of a user is unique (in the group) in the sense that, x-coordinates of the points on $E(F_p)$, representing the group keys of any two users, are different. For example, if $K_i^g = (x_i, y_i)$ and $K_j^g = (x_j, y_j)$ then, $x_i \neq x_j$.

(b) *Group key computation:* Using $rtok_i$, user $U_i$ as well as the server computes the group key $K_i^g$ of $U_i$ as,

$$K_i^g = rtok_i \ . \ G$$

$U_i$ stores $rtok_i$ and $K_i^g$, and the server stores $K_i^g$ as long as $U_i$ is a member of this group.

## 3.3   Secret Share Distribution

Here we will describe a single round of secret share distribution which can be divided into 2 broad steps:

1. Round key computations.

2. Share computation and delegation.

### 3.3.1   Round Key Computations

After the initial steps are over, the server starts the rounds of secret sharing and reconstruction. For each round $r$ of secret share distribution, the server broadcasts its public round key $P_s^r$ into the hierarchy.

$$P_s^r = s_s.G$$

where $s_s$ is a secret integer selected randomly by the server.

The server's public key is used by each user $U_i$ to compute his own round key $K_i^r$, for that round $r$ as

$$K_i^r = rtok_i.P_s^r = (X_i.Y_i).$$

Simultaneously the server can compute the round key of each user $U_i$ using the secret $s_s$ and $U_i$'s group key $K_i^g$ as

$$K_i^r = s_s.K_i^g = (X_i.Y_i).$$

### 3.3.2   Share Computation and Delegation

Now the server starts receiving request messages from the users for secret shares distribution. The request message ($ReqM_i$) of user $U_i$ consists of three pieces of information viz. $[i, parent_i, numberOfChildren_i]$, representing the id of the user ($U_i$) itself, id of its immediate predecessor or parent, and number of its immediate successors or children respectively, in the hierarchy. The server computes and delegates the shares into the highest level (*level 1*) first, followed by *level 2* and so on, upto the lowest level of the hierarchy.

The procedure for share computation and delegation is presented next in form of an algorithm.

**Algorithm 1: Procedure for share computation and delegation**

1: The server computes $threshold_{root} = \lceil TF.(\text{number of } level\ 1 \text{ users})\rceil$, and selects a random $(threshold_{root} - 1)$-degree polynomial $q_{root}$ such that its free coefficient $q_{root}(0) = D$, the original secret data to be shared.
2: **for** (each user $U_i$) **do**
3:     The server receives $ReqM_i$ from $U_i$.
4:     **if** ($numberOfChildren_i == 0$) **then**
5:         The server computes secret share of $U_i$ as $D_i = q_{parent_i}(i)$.
6:     **else**
7:         The server computes $q_{parent_i}(i)$ and splits it into two parts $D_i$ and $D_i'$ using *split operation*. $D_i$ represents the secret share of $U_i$ and $D_i'$ is stored at the server (to be used later, while computing secret shares of $U_i$'s children).
8:         The server computes $threshold_i = \lceil TF\ .\ numberOfChildren_i\rceil$ and selects a random $(threshold_i - 1)$-degree polynomial $q_i$, whose free coefficient $q_i(0) = D_{parent_i}'$.
9:     **end if**
10:    The server sends $D_i$ to $U_i$ over a secure channel (for example , by encrypting with $U_i$'s public key) so that it can be decrypted correctly only by $U_i$.
11: **end for**

### 3.4   Secret Reconstruction

As described in the previous section 3.3.2, the secret share of a user is split into two parts, one of which is held by the user and the other is distributed among its children. Using this concept, secret reconstruction starts from the lowest level and goes up to the higher levels until the root is reached. (Highest level corresponds to level 0; any *level i+1* is lower than *level i*). Each user (except any lowest level user) reconstructs its own secret share with the help of its children and sends it to its parent.

The following 3 steps present the whole reconstruction procedure.

1. All children of $U_i$, co-operate to reconstruct the information $D_i'$ and send it to $U_i$, again over a secure channel.

2. $U_i$ adds $D_i'$ with $D_i$ to produce $q_{parent_i}(i)$, and follows the same step 1 to reconstruct $D_{parent_i}'$ (with the co-operation of its sibling nodes).

3. Steps 1-2 are repeated for nodes at each level, starting from the lowest, moving gradually up to the highest level, when, finally the original secret $D$ is reconstructed at the root server.

### 3.5   Leaving the Hierarchy by a User

When a group member wants to leave the hierarchy, he broadcasts a *leave* message. This message implies that no user in the subtree rooted at the leaving member can participate in secret sharing any more. As soon as this message is received at the server, the server drops the group key of the leaving member. For example, consider the hierarchy shown in figure 2. When $U_2$ leaves the group, none of the users $U_7$ - $U_{14}$ (shown by the shaded circles) can participate in secret sharing, since they form the subtree rooted at $U_1$.

All the users in the hierarchy forming the subtree which was rooted at the leaving member can again participate in secret distribution and reconstruction only after someone joins the hierarchy replacing the leaving member, in other words, when some other user joins the hierarchy in the position that was enjoyed previously by the member who has left.

### 3.6   Our achievements till now

Here we shall try to find out, how far we have succeeded till now, to achieve the goals we had set at the beginning of this paper.

- In our scheme, each user has to store only one secret share, eliminating the storage problem of Shamir's hierarchical secret sharing scheme.

- None of the operations reduces the size of the finite field. In fact, the size of the finite field remains the same, as the one used in general secret sharing schemes without any hierarchy. Thus we have overcome the limitation of Tassa's scheme.

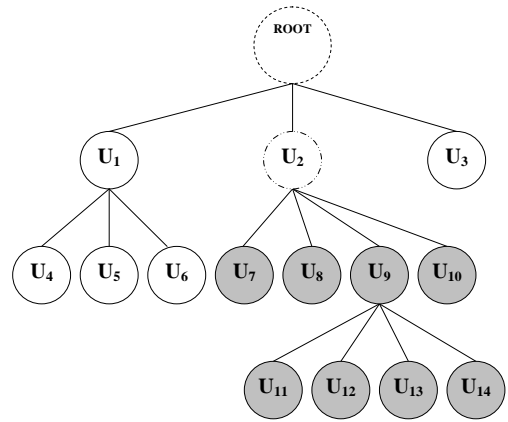

Figure 2: Leave Operation.

So our first two goals have been achieved, by the proposed scheme. Our next concern is the fact, that the security of traditional $(k + 1, n)$-threshold secret sharing depends on the assumption that at no instance of time, more than $k$ shares can be gained by an adversary. This, in many cases proves to be a weak assumption. Next we elaborate on this threat and present a proactive share renewal mechanism to do away with concerned vulnerability.

# 4  Proactive Renewal of Secret Shares in a Hierarchy

Our next concern is the fact, that the security of traditional $(k + 1, n)$-threshold secret sharing depends on the assumption that at no instance of time, more than $k$ shares can be gained by an adversary. This, in many cases proves to be a weak assumption. In this section we elaborate on this threat and present a proactive share renewal mechanism to do away with concerned vulnerability.

## 4.1  Security Model

We consider the case, where the entire lifetime of a $(k+1, n)$-threshold secret sharing scheme, spans a long period of time. In other words, a secret having been shared into a network or group, remains so for a long time period, which may even be a few years. Under such circumstances if an adversary is mobile, i.e. it can hop from one host to another within the network, and its speed is such that it can cover more than $k$ hosts within a lifetime of the secret sharing scheme, the adversary achieves more than $k$ shares and secret is revealed to the adversary. *Thus, we can state that, when an adversary compromises a node of the hierarchy, his aim is to gain the secret share of that node, and over the entire lifetime of the $(k+1, n)-secret\ sharing$ scheme, the aim of the adversary is to gain more than $k$ secret shares.*

If the network is huge, it may infeasible for one adversary to cover more than $k$ hosts within one lifetime of the secret sharing scheme. But here we deal with a hierarchy which exists within an organization and in most cases such a hierarchy consists of a rather small number of users and we use a $(k+1, n)$-threshold secret sharing scheme. So, it is necessary to provide mechanisms so that even a fast mobile adversary cannot achieve more than $k$ shares, however long the lifetime of the scheme be.

So we propose a mechanism to renew the shares within a hierarchy. The proposed mechanism depends on the research work done by Herzberg et. al.[3] on proactive secret sharing. We propose to divide the entire lifetime of the the hierarchical secret sharing scheme into a number of small time periods (so called *proactive*), which may be one day or one week, depending on the speed of the adversary. Secret shares are renewed during each time period.

Even though the proposed proactive share renewal scheme prevents an adversary to compromise more than $k$ users, but the adversary can always compromise $k$ or lesser users, in a single time period. In this paper we propose a mechanism, using elliptic curve cryptography, to detect such compromised users within the hierarchy. But, what does the system do after it has detected a user to be compromised? Here our first assumption comes into play.

**Assumption 1.** When the scheme detects the system of a user to have been compromised by an adversary during a time period, the user sends an alert message to the system administrator who takes necessary measures (such as reboot or running an anti-virus program) so that the user gets rid of the adversary within that time period.

## 4.2 Synchronization

For this scheme to work efficiently, it is necessary that all users of the scheme renew their shares in each time period, without failure.

One way of achieving this is by synchronizing the local clocks of all users with a common global clock. With each tick of the global clock, one time period ends and another starts, alerting all users so that they start another round of share renewal.

But complexities arise, in distributed environments, while trying to syn-chronize large number of users with a single clock since it is very difficult to maintain a single (global) clock, so that all nodes of the hierarchy are synchronized with that clock.

To reduce such complexities, we restrict the notion of synchronization by our second assumption, so that the need to maintain a system-wide global clock, is eliminated, but at the same time, the requirement of dividing the lifetime of the secret sharing scheme into timeperiods as well as periodic share renewal at each node is achieved.

**Assumption 2.** We assume that each node of the tree is synchronized with its parent. Thus, it is sufficient to synchronize all other nodes in a 2-leveled subtree (a $m$-leveled tree is a tree consisting of exactly $m$ levels), with the root of the subtree, thereby eliminating the need for synchronizing all users in the hierarchy.

By saying, *'all other nodes in a 2-leveled subtree are synchronized with the root of the subtree'*, we mean, any message broadcasted, unicasted or multicasted by the root, is received by the other node(s) or recipient(s) within one time period, which is necessarily that time period, in which the message was sent.

### 4.3 Share Renewal Protocol in a 2-leveled Subtree

The secret share held by a user $U_i$ in a 2-leveled subtree, during a time period $t$, is represented by $q(i)^t$. Unique identity of user $U_i$ is represented by $i$ here. The share renewal protocol is carried out in following 3 steps:

1. At the end of time period $t$ the root $U_i$ of the subtree selects a random $k$-degree polynomial $\Delta_i^{t+1}$ where $\Delta_i^{t+1}(0)=0$, i.e. $\Delta_i^{t+1}(x) = \sum_{h=1}^{k} \Delta_{ih}^{t+1}.x^h$ for any $x$. (Here we consider the case of a $(k+1, n)$-threshold secret sharing).

2. $U_i$ sends $\Delta_i^{t+1}(j)$ to each of its children $U_j$ over a secure channel. (For example, by encrypting with $U_j$'s public key. So it can be decrypted by no one but $U_j$).

3. Each child $U_j$ of root $U_i$ renews his share $q(j)^{t+1} = q(j)^t + \Delta_i^{t+1}(j)$.

During each time period, each node verifies whether the $\Delta$ value sent to it by its parent is correct, in order to detect whether the parent is compromised by an adversary. To facilitate this feature, we propose that each user carries out a detection mechanism after step 1 of the share renewal scheme which we describe next.

### 4.4 Detecting a Compromised Node

This section provides a mechanism to detect whether any node has been compromised by adversary. This mechanism is carried out by the nodes in each time period. At the beginning of the detection scheme, the trusted root server selects an elliptic curve $E(F_p)$ on a finite field of order $p$ and makes a base point $G$ on $E(F_p)$ publicly available to all the users in the hierarchy.

1. In each time period the root $U_i$ of each 2-leveled subtree, multicasts the set $\{\Theta_{ih}$ for $1 \leq h \leq k : \Theta_{ih} = \Delta_{ih}.G$ and $\Delta_i(x) = \sum_{h=1}^{k} \Delta_{ih}.x^h\}$ to its children. ($\Delta_{i0}$ or $\Delta_i(0)$ is always 0.)

2. Each user $U_j$ verifies whether the $\Delta$ value sent by its parent $U_i$ is correct by evaluating the following equation:
   $\Delta_i(j).G = \sum_{h=1}^{k} \Theta_{ih}.j^h$.

3. If the equation holds then $U_j$ decides that $U_i$ is not compromised and renews its share using $\Delta_i(j)$.

4. If the equation does not hold $U_j$ decides that $U_i$ is compromised and sends an alert to the system administrator claiming $U_i$ is compromised.

When $U_j$ claims $U_i$ to be compromised, the probability for $U_i$ to be actually compromised is the same as the probability of $U_j$ being compromised, since an adversary may always make a false claim. We propose to resolve this issue in the following way.

### 4.4.1 Deciding on a Claim

In $(k + 1, n)$-threshold proactive secret sharing at most $k$ locations can be compromised by adversary, in one time period. Conversely, at least $(n - k)$ users can be trusted in one time period. Depending on those $(n - k)$ trusted nodes, decision is made in our scheme regarding whether it is the claimed node or the claimer node which is compromised. If $(n - k)$ or more claims are made against a parent node by its children, then it is the claimed node who is decided to be compromised. If, on the other hand, less than $(n - k)$ claims are made against a node, then all of the claimer nodes are decided to be compromised.

Proposed way of finding the compromised node(s) in a claim uses the fact that, the decision of all the non-compromised children of a node, is always the same, regarding whether their parent is compromised or not.

The proactive share renewal mechanism achieves our third goal, since, now we can claim that even if a mobile adversary can obtain more than $k$ secret shares throughout the whole lifetime of the secret sharing scheme, the number of these shares corresponding to a single time period is always less than $k$. In the next section, we do a security analysis and show how integrity, availability and confidentiality are maintained in our scheme.

# 5 Analysis

In this section we will discuss our achievements, to see how far we have succeeded in achieving our goals set at the beginning of this paper.

## 5.1 Analysis of Hierarchical Secret Sharing Scheme: A Comparison with Existing Schemes

The first goal set by us, was to propose a hierarchical secret sharing scheme that will overcome the limitations of the two most well-known and extensively used schemes proposed by Shamir [1] and Tassa [2]. In section **??** we have proposed a scheme that shares a secret data into a hierarchy. Next we shall try to find out how the proposed scheme overcomes the limitations of Shamir's [1] and Tassa's [2] schemes.

- In our scheme, secret share of each parent node is split into two parts. One of those represents the free co-efficient a polynomial used to compute secret shares of its children. The share received by each child node, is the value of the polynomial, computed using one part of the parent's share and the unique registration token of the child.

  Thus each user has to store only one secret share, eliminating the storage problem of Shamir's [1] hierarchical secret sharing scheme.

- Moreover, the split operation does not reduce the size of the finite field, when we go down to the lower layers of the hierarchy implementing our scheme. In fact, the size of the finite field remains the same as the one used in general secret sharing schemes without any hierarchy. Thus we have overcome the limitation of Tassa's [2] scheme.

## 5.2 Analysis of the Proactive Share Renewal Scheme

The major goal that we have tried to achieve in section **??**, is to make our hierarchical secret sharing scheme secure, in an environment where mobile adversaries may be present. We have achieved the same by disallowing such an adversary to acquire more than $k$ shares, in a hierarchical $(k + 1,n)$-threshold secret sharing environment.

Our scheme is based on the proactive secret sharing scheme of Herzberg et. al. [3]. Our scheme has increased the applicability of the Herzberg et. al.'s [3] scheme to a large hierarchy considerably. Our achievements in comparison with Herzberg et. al.'s [3] work are presented in the next section.

### 5.2.1 Comparison with Herzberg et. al.'s [3] Proactive Secret Sharing Scheme

Time Complexity Analysis

The proactive share renewal mechanism of Herzberg et. al. [3] proposes to renew the share of each user in a group, by adding to its own share, a polynomial sent by each of the other users. The time complexity of share renewal by each user in each time period is $O(n)$, where $n$ is the number of users. Thus the time complexity of share renewal during one time period is $O(n^2)$.

Our scheme proposes to renew the share of each user, by adding to it a polynomial sent by its parent. All users do not need to contribute to the share renewal of each of the other users, which makes our scheme applicable to a hierarchy with various levels. Moreover the time complexity of share renewal, during one time period, is reduced to $O(n)$, since each user, during a particular time period, requires $\Theta(1)$ time.

### Detecting a Compromised Node

In section 4.4 we have used elliptic curve cryptography to detect a compromised node, which makes our scheme faster than that of Herzberg et. al. [3].

### Resolving Accusations

A significant part of any proactive secret sharing scheme is to detect whether a node has been compromised by an adversary. And if one node finds another one to be compromised, it accuses the later. But since in such an environment both the accused node as well as the accusing node may be compromised, it is necessary to decide on this matter.

In Herzberg at. al.'s scheme [3], if a node $P_i$ accuses node $P_j$, the later one has to defend itself publicly by broadcasting some message(s), thereby revealing some information (as described in 2.3). We solve the accusation problem in our scheme in the following way (section **??**):

If at least $(n-k)$ out of $n$ children of a node claim it to be compromised, then the parent is compromised indeed. Otherwise it is the claiming child node, which is compromised. Thus we do away with leakage of information as well as the broadcasting requirement, thereby reducing the bandwidth requirement.

## 5.2.2   Security Analysis

### Integrity and Availability

The main objective of proactive secret sharing is to renew the share of each user, in each time period, without modifying the original secret that is shared. By our hierarchical secret sharing scheme, any secret data that is shared among a group of users is represented by the free coefficient of a polynomial (which is essentially a feature of traditional

Shamir's [1] secret sharing scheme). So, the integrity of this free co-efficient needs to be maintained even though the polynomial or the shares are renewed.

In our proactive share renewal scheme, we add a random polynomial $\Delta(i)$ to $q(i)$, the secret share of user $U_i$, in order to renew the share. Since $\Delta(0)$ is always chosen to be 0, $q(0)$ always remains unaltered after the addition operations, although all other coefficients of $q(i)$ are refreshed. Thus we renew the secret share $q(i)$ without modifying the original secret data $q(0)$. The secret data can always be generated by the reconstruction procedure described in section 3.4 and the reconstruction always generates the same secret, in whichever time period it is carried out.

Confidentiality

In the compromised node detection scheme, user $U_i$ matches $\Delta_i(j).G$ value with $\sum_{h=1}^{k} \Theta_{ih}.j^h$, for $1 \leq h \leq k$, where $U_j$ is the parent of $U_i$. We claim that, if both $U_i$ and $U_j$ are non-compromised, then these two values should match. The correctness of our claim is proved, since:
$\Delta_i(j).G$
$= (\Delta_{i1}.j^1 + \Delta_{i2}.j^2 + ... + \Delta_{ik}.j^k).G$ [Note, $\Delta_{ik}$ is always 0.]
$= \Delta_{i1}.G.j^1 + \Delta_{i2}.G.j^2 + ... + \Delta_{ik}.G.j^k$
$= \Theta_{i1}.j^1 + \Theta_{i2}.j^2 + ... + \Theta_{ik}.j^k$
$= \sum_{h=1}^{k} \Theta_{ih}.j^h$

Although the $\Theta$ values (which are obtained by elliptic curve point multiplication of $\Delta$ values and $G$), are multicast into the 2-leveled subtrees, the $\Delta$ values can never be gained by an adversary due to the hardness of Elliptic curve discrete logarithm problem. Thus confidentiality of the renewed shares is always maintained.

# 6 Conclusion

An efficient secret sharing scheme, suitable for a hierarchy of users has been proposed in this paper. Elliptic curve cryptography has been used to generate the keys. Hence the keys are of smaller size and the computations are faster and less complex. The scheme reduces the storage requirement of the users and works without decreasing the size of finite field.

This paper also presents an efficient proactive share renewal mechanism, which removes the vulnerabilities of the hierarchical secret sharing scheme, due to the presence of a mobile adversary within a hierarchy of users. Since the adversary is mobile it compromises some of the users. A protocol to detect such compromised users has been presented, which is based on elliptic curve cryptography. In our scheme we propose that the system administrator takes the necessary steps after detecting any compromised node.

# References

[1] Adi Shamir. (1979). How to share a secret. Communications of the ACM, 22(11), pp. 612-613.

[2] Tamir Tassa. (2007). Hierarchical Threshold Secret Sharing. Journal of Cryptology, 20(2), pp.237-264.

[3] A. Herzberg, S. Jarecki, H. Krawczyk, M. Yung. (1995). Proactive Secret Sharing or: How to cope with perpetual leakage. CRYPTO 1995, LNCS 963, Springer-Verlag 1995, pp. 339-352.

[4] Ed Dawson and Diane Donovan. (1994). The breadth of Shamirs secret sharing scheme. Computers and Security 13, pp. 69-78.

[5] R. Otrovsky, M. Yung (1991). How to withstand mobile virus attacks. Proceedings of the 10th ACM Symposium on the Principles of Distributed Computing, pp.51-61.

[6] D. Schultz. (2007). Mobile proactive secret sharing. In Master's Thesis, MIT.

[7] L. Zhou, F.B. Schneider, R. Van Renesse. (2005). APSS: Proactive secret sharing in asynchronous systems. TISSEC 8(3), pp. 259-286.

[8] D.R. Stinson. (2002). Cryptography: Theory and practice. CRC Press.

[9] Darrel Hankerson, Alfred Menezes, Scott Vanstone. (2004). Guide to Elliptic Curve Cryptography. Springer.

[10] SEC 1: Elliptic Curve Cryptography. (August 22, 2008). Standards for Efficient Cryptography 1 (SEC1). Version 1.9. Working Draft.

[11] Standards for Efficient Cryptography. (2000). White Paper, Certicom Research.

[12] Fuh-Gwo Jeng, Chung-Ming Wang. (2006). An Efficient Key-Management Scheme for Hierarchical Access Control Based on Elliptic Curve Cryptosystem. The Journal of Systems and Software 79, pp. 1161-1167.

[13] Yu Fang Chung, Hsiu Hui Lee, Feipei Lai, Tzer Shyong Chen. (2008). Access Control in User Hierarchy based on Elliptic Curve Cryptosystem. Information Sciences 178, pp. 230-243.

[14] Chris Charnes, Keith Martin, Josef Pieprzyk, Rei Safavi-Naini. (1997). Secret Sharing in Hierarchical Groups. Proceedings of the First International Conference on Information and Communication Security, LNCS 1334, pp. 81-86.

[15] Chin-Chen Chang, Chu-Hsing Lin, Wei Lee, Pai-Cheng Hwang. (2004). Secret Sharing with Access Structures in a Hierarchy. Proceedings of the 18th International Conference on Advanced Information Networking and Application 2 (AINA04), pp. 31.

[16] Hossein Ghodosi, Josef Pieprzyk, Rei Safavi. (1998). Secret Sharing in Multilevel and Compartmented Groups. Australasian Conference on Information Security and Privacy (ACISP'98), LNCS 1438, pp. 367-378.

[17] Giuseppe Ateniese, Alfredo De Santis, Anna Lisa Ferrara, Barbara Masucci. (2006). Provably-Secure Time-Bound Hierarchical Key Assignment Schemes. Proceedings of the 13th ACM conference on Computer and communications security, Conference on Computer and Communications Security, pp. 288-297.

[18] J. Benaloh, J. Leichter. (1990). Generalized secret sharing and monotone functions. Advances in cryptology-CRYPTO88, LNCS, pp. 27-35.

[19] Kamer Kaya and Ali Aydn Selcuk. (2008). A Verifiable Secret Sharing Scheme Based on the Chinese Remainder Theorem. INDOCRYPT 2008, LNCS 5365, pp. 414-425, 2008.

[20] Josef Pieprzyk and Xian-Mo Zhang. (2004). On Cheating Immune Secret Sharing. Discrete Mathematics and Theoretical Computer Science 6, pp. 253-264.

[21] Zhifang Zhang, Mulan Liu, Yeow Meng Chee, San Ling and Huaxiong Wang. (2008). Strongly Multiplicative and 3-Multiplicative Linear Secret Sharing Schemes. ASIACRYPT 2008, LNCS 5350, pp. 19-36.

[22] Massoud Hadian Dehkordi, Samaneh Mashhadi. (2008). Verifiable secret sharing schemes based on non-homogeneous linear recursions and elliptic curves. Computer Communications 31, pp. 1777-1784.

[23] Ronald L. Rivest, Adi Shamir, and Yael Tauman. How to Leak a Secret. pp. 554-567. 1978.

[24] Ronald Cramer, Yevgeniy Dodis, Serge Fehr, Carles Padro, and Daniel Wichs. Detection of Algebraic Manipulation with Applications to Robust Secret Sharing and Fuzzy Extractors. EUROCRYPT 2008, LNCS 4965, pp. 471-488, 2008.